\documentclass[12pt]{article}

\usepackage{graphicx}

\def\fnote#1#2{\begingroup\def\thefootnote{#1}\footnote{#2}\addtocounter
{footnote}{-1}\endgroup}

\begin{document}

\hfill{UTTG-09-06}

\vspace{36pt}

\begin{center}
{\large {\bf {A No-Truncation Approach to Cosmic Microwave Background Anisotropies}}}

\vspace{36pt}
Steven Weinberg\fnote{*}{Electronic address:
weinberg@physics.utexas.edu}\\
{\em Theory Group, Department of Physics, University of
Texas\\
Austin, TX, 78712}

\vspace{30pt}

\noindent
{\bf Abstract}
\end{center}
\noindent
We offer a method of calculating the source term in the line-of-sight integral for cosmic microwave background anisotropies without using a truncated partial-wave expansion in the Boltzmann hierarchy.
 \vfill

\pagebreak

\def\BM#1{\mbox{\boldmath{$#1$}}}

\begin{center}
{\bf I. Introduction}
\end{center}

Originally the Boltzmann  equation for the photon distribution in cosmology[1] was solved numerically by expanding the components of the photon density matrix in a series of Legendre polynomials[2], but to get results that could be compared with observation of the cosmic microwave background this method requires the inclusion of hundreds or even thousands of partial waves, requiring hours or even days of computer time for each theoretical model.   A great improvement was introduced with the suggestion to use instead a formal solution of the Boltzmann equation,  in the form of a ``line of sight'' integral[3].  But this is still only  a formal solution, in the sense that we still need to calculate  source terms appearing in the integrand.  These terms involve partial waves for the photon distribution up to $\ell=2$ for the scalar modes and $\ell=4$ for the tensor modes, and these of course are coupled to higher partial waves.  In  the original proposal of the ``line of sight'' method, and in the computer programs CMBfast and CAMB based on this method, these source terms are calculated numerically, by first finding an approximate solution of the Boltzmann equations for partial wave amplitudes.  An accurate solution for the  partial waves appearing in the source terms can be found by truncating the partial wave expansion at a sufficiently high value of $\ell$.  In the latest version of 
CMBfast, the integrand for scalar modes is calculated using partial waves up to   $\ell=12$, in which case one has to solve at least 26  coupled ordinary differential equations for the evolution of the partial wave amplitudes, not counting the equations needed to follow the evolution of the baryonic plasma, cold dark matter, neutrinos, and gravitational field components.  For tensor modes, the source terms are calculated by solving the 22 differential equations for partial waves with up to $\ell=10$ for photons.    The results for $\ell\leq 2$ or $\ell\leq 4$ are used in this method to calculate the integrand of the line-of-sight integral, which then is used to calculate all the higher partial wave amplitudes measured in observations of the cosmic microwave background, up to values of $\ell$ over 1000. 

This article will present an alternative approach, which does not use  partial wave expansions to calculate the source terms, and hence obviates the need for any truncation of this expansion.  Instead of a large number of coupled differential equations for the partial waves, we have a single integral equation for the tensor modes, and a trio of coupled integral equations for the scalar modes (including one for the plasma velocity).  Of course, integral equations are generally harder to solve numerically than differential equations (no routine for solving them is supplied by Mathematica), but in the case at hand the integral equations can be solved numerically by simple iteration.  In this method, the calculation itself provides  an immediate way of judging its own  reliability --- if the $n$th iteration agrees with the $n-1$th iteration to a satisfactory degree of accuracy, one has a solution.  In a sample calculation of the source term for the tensor modes,  the results converge rapidly in just a few iterations.

This paper concentrates in the next  section on the calculation of the photon distribution, but a truncated partial wave expansion is also unnecessary for neutrinos.  Indeed, it is already known[4] that the momentum distribution of massless neutrinos for a given metric perturbation can be calculated in terms of a simple line-of-sight integral, with no need to solve integral equations.  CMBfast does not use this line-of-sight method for neutrinos, presumably because no one is interested in very high partial waves in the neutrino distribution, but to get good accuracy for the neutrino contribution to the energy-momentum tensor it carries the partial wave expansion to $\ell_{\rm max}=25$.  In the last section of this paper the approach of reference [4], which dispenses with partial wave expansions, is extended to the case of massive as well as massless neutrinos, and to scalar as well as tensor modes.

\begin{center}
{\bf II. Photons}
\end{center}

First, some reminders about the Boltzmann equation for  photons in cosmology.  We will adopt a coordinate system in which the metric takes the form
\begin{equation}
g_{00}({\bf x},t)=-1\;,~~~g_{0i}({\bf x},t)=0\;,~~~g_{ij}({\bf x},t)=a^2(t)\,\Big(\delta_{ij}+h_{ij}({\bf x},t)\Big)\;,
\end{equation}
where $h_{ij}$ is a first-order  perturbation.  Weakly perturbed metrics will automatically be of this form in tensor modes, and can be put in this form for scalar modes by adopting a synchronous gauge.  

For our purposes, it is important to write the Boltzmann equation for the photon distribution in a matrix form, rather than in the partial wave formalism in which it is usually presented.   The photon distribution is described by a polarization density matrix $n^{ij}({\bf x},{\bf p},t)$, defined so that if we measure whether photons have polarization in a direction $e^i$ rather than in an orthogonal direction, then the number of photons with polarization $e^i$ in a volume $\prod_i dp_i\,dx^i$ of phase space at time $t$ will be found to be  $g_{ik}g_{jl}e^ke^ln^{ij}({\bf x},{\bf p},t)\prod_m dp_m\,dx^m$, with $p_i n^{ij}=0$.  (The polarization of a photon with 3-momentum $p_i$ is described by a polarization vector $e^i$, satisfying $p_ie^i=0$ and $g_{ij}e^ie^j=1$.)  For small perturbations, this matrix can be put in  the form
\begin{eqnarray}
n^{ij}({\bf x}, {\bf p}, t)&=&\frac{1}{2}\bar{n}_\gamma\left(a(t)\sqrt{g^{kl}({\bf x},t)p_kp_l}\right)\left[g^{ij}({\bf x},t)-\frac{g^{ik}({\bf x},  t)g^{jl}({\bf x}, t)p_kp_l}{ g^{kl}({\bf x},t)p_kp_l }\right]\nonumber\\&&~~~~~~
+\delta n^{ij}({\bf x}, {\bf p}, t)\;.
\end{eqnarray}
Here $\bar{n}_\gamma(p)$ is the equilibrium phase space number density\begin{equation}
\bar{n}_\gamma(p)\equiv \frac{1}{(2\pi)^3}\left[\exp\Big(p/k_{\cal B}a(t)T(t)\Big)-1\right]^{-1}\;,
\end{equation}
(which is a time-independent function of its argument because in the era of interest $T(t)\propto 1/a(t)$), and $\delta n^{ij}$ is a small perturbation.  This perturbation satisfies a linearized Boltzmann equation:
\begin{eqnarray}
&&\frac{\partial\,\delta n^{ij}({\bf x},{\bf p},t)}{\partial t}
+\frac{\hat{p}_k}{a(t)}\frac{\partial\,\delta n^{ij}({\bf x},{\bf p},t)}{\partial x^k}
+\frac{2\dot{a}(t)}{a(t)}\,\delta n^{ij}({\bf x},{\bf p},t)
\nonumber\\&&~~~~~-\frac{1}{4a^2(t)}p\bar{n}'_\gamma(p)\hat{p}_k\hat{p}_l\dot{h}_{kl}({\bf x},t)\,\Big(\delta_{ij}-\hat{p}_i\hat{p}_j\Big)\nonumber\\&&=-\omega_c(t)\,\delta n^{ij}({\bf x},{\bf p},t)+\frac{3\omega_c(t)}{8\pi}\int d^2\hat{p}_1 \nonumber\\&&~~~~\times\Big[\delta n^{ij}({\bf x},p\hat{p}_1,t)-\hat{p}_i\hat{p}_k \,\delta n^{kj}({\bf x},p\hat{p}_1,t)-\hat{p}_j\hat{p}_k\,\delta n^{ik}({\bf x},p\hat{p}_1,t)\nonumber\\&&~~~~~+\hat{p}_i\hat{p}_j\hat{p}_k\hat{p}_l\,\delta n^{kl}({\bf x},p\hat{p}_1,t)\Big]\nonumber\\&&~~~-\frac{\omega_c(t)}{2a^2(t)}\,p_k \delta u_k({\bf x},t)\,\bar{n}'_\gamma(p)\Big[\delta_{ij}-\hat{p}_i\hat{p}_j\Big]
\;,
\end{eqnarray}
where $p\equiv \sqrt{p_ip_i}$, $\hat{p}_k\equiv p_k/p$, $\delta u_k$ is the streaming velocity of the baryonic plasma, and $\omega_c$ is the  frequency with which a photon collides with electrons in the plasma.  Instead of $\delta n^{ij}$, it is sufficient to consider the intensity matrix perturbation
$J_{ij}({\bf x},\hat{p},t)$, defined by
\begin{equation}
a^4(t)\,\bar{\rho}_\gamma(t)\, J_{ij}({\bf x},\hat{p},t)\equiv a^2(t)\,\int_0^\infty \delta n^{ij}({\bf x},p\hat{p},t)\,4\pi p^3\,dp\;,
\end{equation}
where $\bar{\rho}_\gamma(t)\equiv a^{-4}(t)\int 4\pi p^3\bar{n}_\gamma(p)\,dp$
is the mean photon energy density.
(This is all we need to calculate the photon contributions to the perturbations in the energy-momentum tensor.)  
To derive the Boltzmann equation for $J_{ij}({\bf x},\hat{p},t)$ we multiply Eq.~(4) with $4\pi p^3 $ and integrate over $p\equiv \sqrt{p_ip_i} $, and find
\begin{eqnarray}
&&\frac{\partial\,J_{ij}({\bf x},\hat{p},t)}{\partial t}
+\frac{\hat{p}_k}{a(t)}\frac{\partial\,J_{ij}({\bf x},\hat{p},t)}{\partial x^k}
\nonumber\\&&~~~~~+\hat{p}_k\hat{p}_l\dot{h}_{kl}({\bf x},t)\,\Big(\delta_{ij}-\hat{p}_i\hat{p}_j\Big)\nonumber\\&&=-\omega_c(t)\,J_{ij}({\bf x},\hat{p},t)+\frac{3\omega_c(t)}{8\pi}\int d^2\hat{p}_1 \nonumber\\&&~~~~\times\Big[J_{ij}({\bf x},\hat{p}_1,t)-\hat{p}_i\hat{p}_k \,J_{kj}({\bf x},\hat{p}_1,t)-\hat{p}_j\hat{p}_k\,J_{ik}({\bf x},\hat{p}_1,t)\nonumber\\&&~~~~~+\hat{p}_i\hat{p}_j\hat{p}_k\hat{p}_l\,J_{kl}({\bf x},\hat{p}_1,t)\Big]\nonumber\\&&~~+2\omega_c(t) \Big[\delta_{ij}-\hat{p}_i\hat{p}_j\Big]\,\hat{p}_k\delta u_k({\bf x},t)
\;.
\end{eqnarray}
We can also calculate the equation of motion of the plasma, using the perturbed momentum-conservation equation:
\begin{equation}
\frac{\bar{\rho}_B(t)}{a(t)}\frac{\partial}{\partial t}\Big[a(t)\delta u_k({\bf x},t)\Big]=-\frac{4}{3}\omega_c(t)\bar{\rho}_\gamma(t)\delta u_k({\bf x},t)+\omega_c(t)\bar{\rho}_\gamma(t)\int \frac{d^2\hat{p}}{4\pi}\;J_{ii}({\bf x},\hat{p},t)\,\hat{p}_k\;,
\end{equation}
where $\bar{\rho}_B$ is the unperturbed density of the baryonic plasma.
These partial differential equations can be converted to ordinary differential equations by writing the perturbed metric as a Fourier integral
\begin{equation}
h_{ij}({\bf x},t)=\int d^3p\; e^{i{\bf q}\cdot{\bf x}}\,h_{ij}({\bf q},t)\;.
\end{equation}
and looking for solutions in the form
\begin{equation}
J_{ij}({\bf x},\hat{p},t)=\int d^3q\;e^{i{\bf q}\cdot{\bf x}}J_{ij}({\bf q},\hat{p},t)\;,~~~~~~~\delta u_i({\bf x},t)=\int d^3q\;e^{i{\bf q}\cdot{\bf x}}\delta u_i({\bf q},t)\;.
\end{equation}
Then Eqs.~(6) and (7) become ordinary differential equations, with $\partial /\partial x^i$ replaced with $iq_i$.  They have a formal solution in the form of a  ``line of sight'' integral[3], which in the general case may be written
\begin{eqnarray}
&&J_{ij}({\bf q},\hat{p},t)=\int_{t_1}^t dt'\;\exp\left(-i{\bf q}\cdot\hat{p}\int_{t'}^t \frac{dt''}{a(t'')} -\int_{t'}^t dt''\;\omega_c(t'')\right)\nonumber\\&&~~~~\times\Bigg[-\hat{p}_k\hat{p}_l\,\Big(\delta_{ij}-\hat{p}_i\hat{p}_j\Big)\,\dot{h}_{kl}({\bf q},t')\nonumber\\&&+\frac{3\omega_c(t')}{2}
\left({\cal J}_{ij}({\bf q},t')-\hat{p}_i\hat{p}_k{\cal J}_{kj}({\bf q},t')
-\hat{p}_j\hat{p}_k{\cal J}_{ik}({\bf q},t')+\hat{p}_i\hat{p}_j\hat{p}_k\hat{p}_l{\cal J}_{kl}({\bf q},t')\right)\nonumber\\&&~~+2\omega_c(t')[\delta_{ij}-\hat{p}_i\hat{p}_j]\,\hat{p}_k\delta u_k({\bf q},t')\Bigg]+J_{ij}({\bf q},\hat{p},t_1)\;,
\end{eqnarray}
\begin{eqnarray}
&&\delta u_i({\bf q},t)=\frac{3}{4a(t)}\int_{t_1}^t dt'\;\exp\left(-\int_{t'}^t dt''\;\frac{\omega_c(t'')}{R(t'')}\right)\frac{\omega_c(t')a(t')}{R(t')}{\cal I}_i({\bf q},t')\nonumber\\&&~~~~~+\delta u_i({\bf q},t_1)\;.
\end{eqnarray}
We have here introduced the convenient abbreviations
\begin{eqnarray}
{\cal J}_{ij}({\bf q},t)&\equiv &\int \frac{d^2p}{4\pi}\,J_{ij}({\bf q},\hat{p},t)\;,\\
{\cal I}_{i}({\bf q},t)&\equiv &\int \frac{d^2p}{4\pi}\,J_{kk}({\bf q},\hat{p},t)\,\hat{p}_i\;,
\end{eqnarray}
and, as usual,
$$
R(t)\equiv\frac{3\bar{\rho}_B(t)}{4\bar{\rho}_\gamma(t)}\;.
$$
If we take the  initial time $t_1$  early enough so that photons are in local thermal equilibrium with the plasma at that time, then 
\begin{equation}
J_{ij}({\bf q},\hat{p},t_1)=2\Big(\delta_{ij}-\hat{p}_i\hat{p}_j\Big)\,\left[\frac{\delta T({\bf q},t_1)}{\bar{T}(t_1)}+\hat{p}_k \delta u_k({\bf q},t_1)\right]\;.
\end{equation}

It is the calculation of the source terms ${\cal J}_{ij}({\bf q},t)$ and ${\cal I}_{i}({\bf q},t)$ that concerns us in this paper.  For this purpose, we now need to distinguish between tensor and scalar modes.  We first consider tensor modes, which are computationally simpler.

\begin{center}
{\bf Tensor Modes}
\end{center}

For tensor modes the metric perturbation takes the form  
\begin{equation}
h_{ij}({\bf q},t)=\sum_{\lambda=\pm 2}\beta({\bf q},\lambda)\,e_{ij}(\hat{q},\lambda)\,{\cal D}_q(t)\;,
\end{equation}
where $\beta({\bf q},\lambda)$ is a stochastic parameter; $e_{ij}(\hat{q},\lambda)$ is a polarization tensor for helicity $\lambda$, with $\hat{q}_i\,e_{ij}({\hat q},\lambda)=0$ and $e_{kk}(\hat{q},\lambda)=0$;  and ${\cal D}_q(t)$ is the  solution of the wave equation 
\begin{equation} 
\ddot{\cal D}_q(t)+3\frac{\dot{a}}{a}\dot{\cal D}_q(t)+
\frac{q^2}{a^2}{\cal D}_q(t)=16\pi G\,\pi^T_q(t)\;,
\end{equation} 
that does not decay while outside the horizon.  Here $\pi^T_q(t)$ is the coefficient of $\sum_\lambda e_{ij}({\bf q},\lambda)\beta({\bf q},\lambda)$ in the Fourier transform of the tensor part of the anisotropic inertia tensor. (We are considering times sufficiently late so that the decaying solution makes a negligible contribution to $\delta g_{ij}$.)
The quantity $J_{ij}({\bf q},t)$ will then take the form of a corresponding sum over graviton helicities
\begin{equation}
J_{ij}({\bf q},\hat{p},t)=\sum_{\lambda =\pm 2}\beta({\bf q},\lambda)\,J_{ij}({\bf q},\hat{p},t,\lambda)\;,
\end{equation}
with 
$J_{ij}({\bf q},t,\lambda)$  ordinary c-number functions, not stochastic fields, satisfying the  Boltzmann equation
\begin{eqnarray}
&&\frac{\partial\,J_{ij}({\bf q},\hat{p},t,\lambda)}{\partial t}
+i\frac{{\bf q}\cdot\hat{p}}{a(t)}J_{ij}({\bf q},\hat{p},t,\lambda)
\nonumber\\&&~~~~~+\hat{p}_k\hat{p}_l\,e_{kl}(\hat{q},\lambda)\,\dot{\cal D}_q(t)\,\Big(\delta_{ij}-\hat{p}_i\hat{p}_j\Big)\nonumber\\&&=-\omega_c(t)\,J_{ij}({\bf q},\hat{p},t,\lambda)+\frac{3\omega_c(t)}{2} \nonumber\\&&~~~~\times\Big[{\cal J}_{ij}({\bf q},t,\lambda)-\hat{p}_i\hat{p}_k \,{\cal J}_{kj}({\bf q},t,\lambda)-\hat{p}_j\hat{p}_k\,{\cal J}_{ik}({\bf q},t,\lambda)\nonumber\\&&~~~~~+\hat{p}_i\hat{p}_j\hat{p}_k\hat{p}_l\,{\cal J}_{kl}({\bf q},t,\lambda)\Big]
\;,
\end{eqnarray}
where
\begin{equation}
{\cal J}_{ij}({\bf q},t,\lambda)\equiv \int \frac{d^2\hat{p}}{4\pi}\,J_{ij}({\bf q},\hat{p},t,\lambda)\;.
\end{equation}
(The velocity perturbation $\delta u_i$ is absent in tensor modes.)  Furthermore, because $J_{ij}({\bf q},t,\lambda)$ for a given helicity $\lambda$ must be a linear combination of 
the polarization tensor components $e_{kl}({\hat q},\lambda)$ with the same $\lambda$, while 
$\hat{q}_k\,e_{kl}({\hat q},\lambda)$ and $e_{kk}(\hat{q},\lambda)$ both vanish, the only possible form of  ${\cal J}_{ij}({\bf q},t,\lambda)$ allowed by rotational invariance is just $e_{ij}({\hat q},\lambda)$ times some function of $q\equiv |{\bf q}|$ and $t$.  This relation is conventionally written
\begin{equation}
{\cal J}_{ij}({\bf q},t,\lambda)=-\frac{2}{3}e_{ij}(\hat{q},\lambda)\,\Psi(q,t)\;.
\end{equation}

To make contact with the notation used in the usual calculation of the source term $\Psi(q,t)$, we note that the intensity matrix perturbation may be written in the  form
\begin{eqnarray}
&&J_{ij}({\bf q},\hat{p},t,\lambda)=\frac{1}{2}\Big(\delta_{ij}-\hat{p}_i\hat{p}_j\Big)\,\hat{p}_k\hat{p}_l\,e_{kl}(\hat{q},\lambda)\Big(
\Delta_T^{(T)}(q, \hat{p}\cdot\hat{q},t)+\Delta_P^{(T)}(q, \hat{p}\cdot\hat{q},t)\Big)\nonumber\\&&
~~~+\Big(e_{ij}(\hat{q},\lambda)-\hat{p}_i\hat{p}_k e_{kj}(\hat{q},\lambda)
-\hat{p}_j\hat{p}_k e_{ik}(\hat{q},\lambda)+
\hat{p}_i\hat{p}_j\hat{p}_k\hat{p}_l\,e_{kl}(\hat{q},\lambda)\Big)\,
\Delta_P^{(T)}(q,\hat{p}\cdot\hat{q},t)\;.\nonumber\\&&{}
\end{eqnarray}
 (Here the superscript $T$ stands for `tensor,' while the subscript $T$ stands for `temperature.'  The coefficients are chosen so that $J_{ii}$ is proportional to $\Delta_T$, and the polarization is proportional to $\Delta_P$.)  A third term proportional to $(\hat{q}_i-\hat{p}_i(\hat{p}\cdot\hat{q}))(\hat{q}_j-\hat{p}_j(\hat{p}\cdot\hat{q}))\hat{p}_k\hat{p}_le_{kl}$ would be allowed by symmetry principles, but is not generated by the Boltzmann equation.  
Using Eq.~(21) in Eq.~(18) yields separate Boltzmann equations for 
$\Delta_T^{(T)}$ and $\Delta_P^{(T)}$:
\begin{equation}
\frac{\partial}{\partial t}\Delta_T^{(T)}(q, \hat{p}\cdot\hat{q},t)+i\,a^{-1}(t)\,{\bf q}\cdot\hat{p}\,\Delta_T^{(T)}(q, \hat{p}\cdot\hat{q},t)=-2\dot{\cal D}_q(t)-\omega_c(t)\,\Delta_T^{(T)}(q, \hat{p}\cdot\hat{q},t)+\omega_c(t)\,\Psi(q,t)\;,
\end{equation}
\begin{equation}
\frac{\partial}{\partial t}\Delta_P^{(T)}(q, \hat{p}\cdot\hat{q},t)+i\,a^{-1}(t)\,{\bf q}\cdot\hat{p}\,\Delta_P^{(T)}(q, \hat{p}\cdot\hat{q},t)=-\omega_c(t)\,\Delta_P^{(T)}(q, \hat{p}\cdot\hat{q},t)-\omega_c(t)\,\Psi(q,t)\;.
\end{equation}
To calculate the source term $\Psi(q,t)$ in terms of partial waves, one first integrates Eq.~(21) over $\hat{p}$, and finds
\begin{eqnarray}
&&\Psi(q,t)=-\frac{3}{2}\int \frac{d^2\hat{p}}{4\pi}\,\Bigg[-\frac{1}{8}\Big(1-(\hat{p}\cdot\hat{q})^2\Big)^2\Delta_T^{(T)}(q, \hat{p}\cdot\hat{q},t) \nonumber\\&&~~~~~~~~+\left((\hat{p}\cdot\hat{q})^2+\frac{1}{8}\Big(1-(\hat{p}\cdot\hat{q})^2\Big)^2\right)\,\Delta_P^{(T)}(q, \hat{p}\cdot\hat{q},t)\Bigg]\;.
\end{eqnarray}
The functions $\Delta_T^{(T)}$ and $\Delta_P^{(T)}$ may be expanded in Legendre polynomials
\begin{eqnarray}
\Delta_T^{(T)}(q, \hat{p}\cdot\hat{q},t)&=&\sum_{\ell=0}^\infty i^{-\ell}(2\ell+1)\,P_\ell(\hat{q}\cdot\hat{p})\,\Delta_{T\ell}^{(T)}(q,t)\,\\
\Delta_P^{(T)}(q, \hat{p}\cdot\hat{q},t)&=&\sum_{\ell=0}^\infty i^{-\ell}(2\ell+1)\,P_\ell(\hat{q}\cdot\hat{p})\,\Delta_{P\ell}^{(T)}(q,t)\;,
\end{eqnarray}
and then Eq.~(24) reads[5]:
\begin{eqnarray}
&&\Psi(q,t)=\frac{1}{10}\Delta^{(T)}_{T0}(q,t)+\frac{1}{7}\Delta^{(T)}_{T2}(q,t)+
\frac{3}{70}\Delta^{(T)}_{T4}(q,t)-\frac{3}{5}\Delta^{(T)}_{P0}(q,t)\nonumber\\&&~~~~~~~+\frac{6}{7}\Delta^{(T)}_{P2}(q,t)-\frac{3}{70}\Delta^{(T)}_{P4}(q,t)\;.
\end{eqnarray}
As already mentioned, experience shows that to accurately calculate the partial wave amplitudes up to $\ell=4$, which appear in Eq.~(27), one needs to solve the Boltzmann equations for the partial wave amplitudes up to  larger values of $\ell$, up to $\ell=10$.  Once $\Psi$ is calculated in this way, we can calculate $J_{ij}({\bf q},\hat{p},t,\lambda)$ for very much higher values of $\ell$ by using the ``line of sight'' integral (10), which for the tensor modes gives 
\begin{eqnarray}
&&J_{ij}({\bf q},\hat{p},t,\lambda)=\int_{t_1}^t dt'\;\exp\left(-i{\bf q}\cdot\hat{p}\int_{t'}^t \frac{dt''}{a(t'')} -\int_{t'}^t dt''\;\omega_c(t'')\right)\nonumber\\&&~~~~\times\Bigg[-\hat{p}_k\hat{p}_l\,\Big(\delta_{ij}-\hat{p}_i\hat{p}_j\Big)\,e_{kl}(\hat{q},\lambda)\,\dot{\cal D}_q(t)\nonumber\\&&-\omega_c(t')\,\Psi(q,t')
\Big(e_{ij}(\hat{q},\lambda)-\hat{p}_i\hat{p}_k e_{kj}(\hat{q},\lambda)
-\hat{p}_j\hat{p}_k e_{ik}(\hat{q},\lambda)+\hat{p}_i\hat{p}_j\hat{p}_k\hat{p}_l e_{kl}(\hat{q},\lambda)\Big)\Bigg]\;.\nonumber\\&&{}
\end{eqnarray}
(In tensor modes there is no perturbation to either the temperature or the velocity of the baryonic plasma, so Eq.~(14) gives the initial value $J_{ij}({\bf q},\hat{p},t_1)=0$ for an initial time $t_1$ taken sufficiently early so that photons are in local thermal equilibrium with the plasma.)

Here we suggest the alternative, of deriving an integral equation for $\Psi(q,t)$ by simply analytically integrating Eq.~(28) over $\hat{p}$.  
Equating the coefficients of $e_{ij}$ on both sides gives the integral equation
\begin{eqnarray}
&&\Psi(q,t)=\frac{3}{2}\int_{t_1}^t dt' \exp\left[-\int_{t'}^{t}\omega_c(t'')\,dt''\right]
\nonumber\\&&~~~\times \Bigg[ -2\dot{\cal D}_q(t')K\left(q\int_{t'}^{t}\frac{dt''}{a(t'')}\right)
+\omega_c(t')F\left(q\int_{t'}^t\frac{dt''}{a(t'')}\right)\,\Psi(q,t')\Bigg]\;.\nonumber\\&&{}
\end{eqnarray}
Here $K(v)$ and $F(v)$ are the functions
\begin{equation}
K(v)\equiv j_2(v)/v^2\;,~~~F(v)\equiv  j_0(v)-2j_1(v)/v+2j_2(v)/v^2\;.
\end{equation}
Integral equations of this sort are harder to solve numerically than differential equations, because in a step-by-step calculation it is necessary to keep track of $\Psi(t')$ for all $t'<t$ in order to calculate $\Psi (t)$.  On the other hand, such equations can also be solved by simple iteration, in which the numerical work is reduced to doing a few integrals.  We calculate the $n$th approximation $ \Psi^{(n)}$ to $\Psi$ by using the previous approximation $\Psi^{(n-1)}$ in the second term in square brackets in Eq.~(29), and start this calculation by taking the lowest approximation $\Psi^{(0)}$ as just Eq.~(29) with the second term in square brackets dropped:
\begin{equation}
\Psi^{(0)}(q,t)=-3\int_{t_1}^t dt' \exp\left[-\int_{t'}^{t}\omega_c(t'')\,dt''\right]\,\dot{\cal D}_q(t')K\left(q\int_{t'}^{t}\frac{dt''}{a(t'')}\right)
\;.
\end{equation}
Using $\Psi^{(0)}$  for $\Psi$ in Eq.~(28) would give precisely the same photon distribution function that we would find if we assumed that photons remained  unpolarized until the last scattering; subsequent iterations take account of the polarization produced in multiple scattering.  When after $n$ iterations we find that $\Psi^{(n)}=\Psi^{(n-1)}$ to an adequate degree of accuracy, we have a solution of Eq.~(29).  This is in contrast with the usual truncation method, in which the accuracy of the calculation for a given maximum multipole order can only be judged by comparing with the results for a higher maximum multipole order.

To test the convergence of this iteration procedure, we adopt the usual $\Lambda$CDM model, with  cosmological parameters
$$\Omega_Bh^2=0.0223\;,~~~\Omega_Mh^2=0.1262\;,~~~h=0.732\;.$$
We calculate the collision rate $\omega_c$ by solving the kinetic equations for hydrogen recombination, with an un-ionized helium fraction $Y=0.26$.  The numerical calculations are done with $q$  chosen so that the physical wave number $ q/a$ comes within the horizon at just the time of matter-radiation equality.  The gravitational wave amplitude ${\cal D}_q(t)$ is calculated ignoring the effect of anisotropic inertia.  Over the interesting time interval  when the matter/radiation density ratio $y$ increases from 2 to 4, which includes the time of recombination, the first iteration $\Psi^{(1)}_q(t)$ has roughly the same time-dependence as $\Psi^{(0)}_q(t)$, but is about 13\% to 33 \% larger.  On the other hand, after five iterations we get a result   $\Psi^{(5)}(t)$ that at worst (at $y\simeq 3$) is within  0.3\% of the previous iteration $\Psi^{(4)}$,  and is much closer at other values of $y$.  As a further illustration of the convergence of the iteration procedure, Figure 1 shows the first few iterations for several wave numbers and a broader range of values of $y$. 

\begin{figure}[h]
\begin{picture}(200,200)
\includegraphics{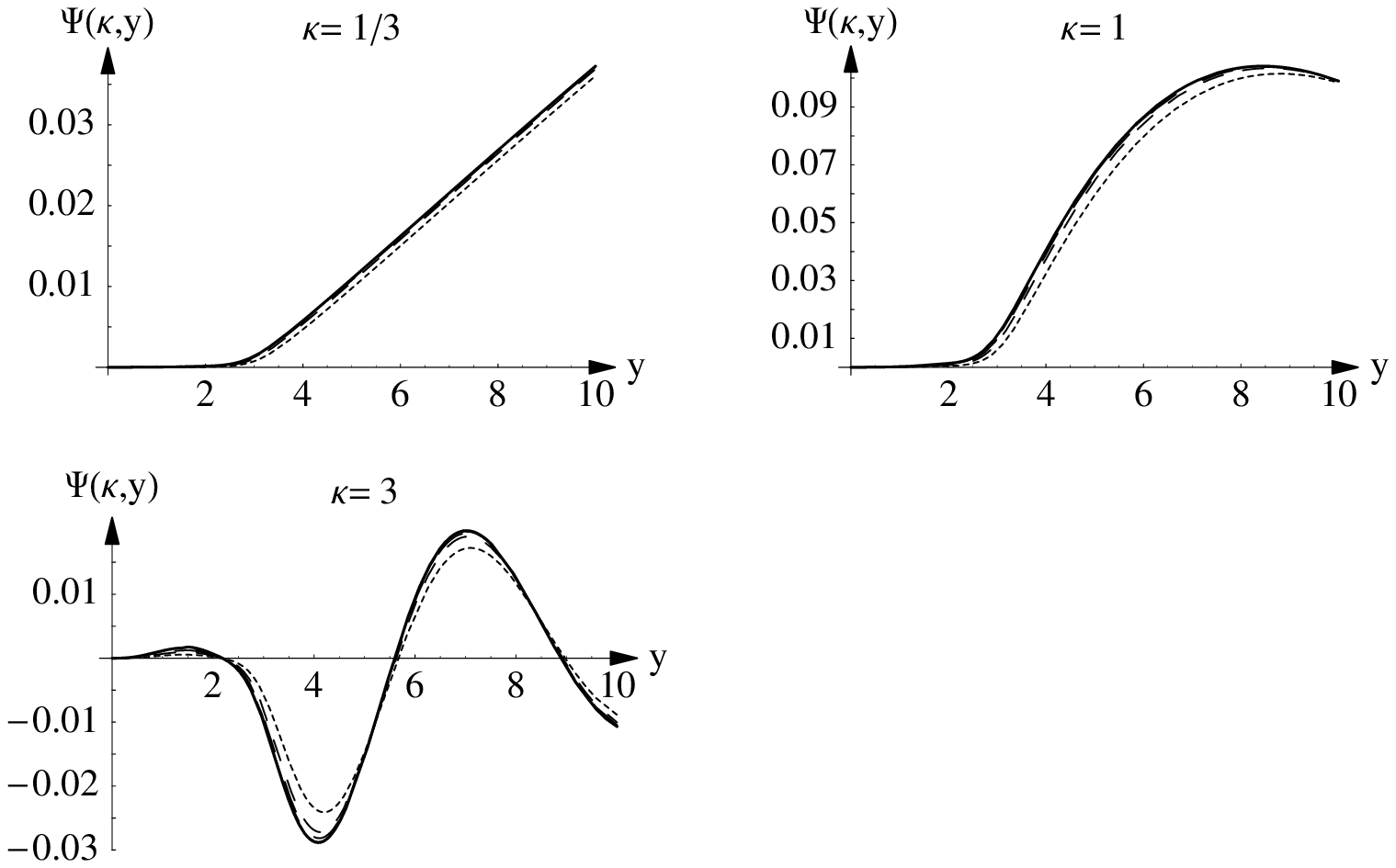}
\end{picture}
\caption{Iterative Solution of Equation (29) for the Source Function $\Psi$.  Short dashes indicate the zeroth iteration (31); longer dashes indicate the first iteration; and further iterations are indistinguishable from the solid curve, which therefore represents the solution.  Here the source function is calculated for a gravitational perturbation ${\cal D}_q(t)$ whose value ${\cal D}_q^o$ before horizon entry is unity; for any other initial condition, $\Psi$ should be multiplied by the value of ${\cal D}_q^o$.  The quantity $\kappa$ is the wave number $q$ in units of the wave number that just comes into the horizon at matter-radiation equality, and $y$  is the Robertson--Walker scale factor, in units of the scale factor at matter-radiation equality.  The calculation was done by R. Flauger using the electron number density calculated with the program Recfast [S. Seager, D. D. Sasselov, and D. Scott, Astrophys. J. {\bf 523}, L1 (1999); Astrophys. J. Suppl. {\bf 128}, 407 (2000)] and taking $\Omega_Mh^2=0.133$, $\Omega_Bh^2=0.02238$, $h=0.72$, $T_0=2.725$K, and $Y_{\rm He}=0.24$. }
\end{figure}

We will not go on here to use this source function in Eq.~(28) to calculate the intensity matrix perturbation, because we are not proposing anything new in that part of the calculation of the microwave background anisotropies, but only in the calculation of the source terms appearing in the integrand of the line of sight integral.  The sample calculation described here shows that this is a practical method of calculating the source terms, as well as having the conceptual advantage of avoiding a more-or-less arbitrary truncation of a partial wave expansion.

\begin{center}
{\bf Scalar Modes}
\end{center}

For scalar modes, the metric perturbation in synchronous gauge takes the form
\begin{equation}
h_{ij}({\bf q},t)=\delta_{ij}A({\bf q},t)-{\bf q}_i{\bf q}_j\,B({\bf q},t)\;.
\end{equation}
Assuming the perturbation to be dominated by a single mode (presumably the  adiabatic mode that does not decay while outside the horizon), the dependence of $A({\bf q},t)$ and $B({\bf q},t)$ on the direction of ${\bf q}$ is entirely contained in a stochastic factor $\beta({\bf q})$ for this mode:
\begin{equation}
A({\bf q},t)=\beta({\bf q})\,A_q(t)\;~~~~~B({\bf q},t)=\beta({\bf q})\,B_q(t)\;.
\end{equation}
In this case, the source terms (12)  and (13) appearing in the integrand of the line-of-sight integral 
must take the form
\begin{equation}
{\cal J}_{ij}({\bf q},t)=\beta({\bf q})\,\left[\delta_{ij}\Phi(q,t)+\frac{1}{2}\hat{q}_i\hat{q}_j\Pi(q,t)\right]\;,
\end{equation}
\begin{equation}
{\cal I}_i({\bf q},t)=i\beta({\bf q})\,\hat{q}_i\,\Omega(q,t)
\end{equation}
Then,  using Eq.~(11) to evaluate the plasma velocity,  the line-of-sight formula (10) becomes
\begin{eqnarray}
&&J_{ij}({\bf q},\hat{p},t)=\beta({\bf q})\,\int_{t_1}^t dt'\;\exp\left(-i{\bf q}\cdot\hat{p}\int_{t'}^t \frac{dt''}{a(t'')} -\int_{t'}^t dt''\;\omega_c(t'')\right)\nonumber\\&&~~~~\times\Bigg[-\Big(\delta_{ij}-\hat{p}_i\hat{p}_j\Big)\,\Big(\dot{A}_q(t')-(\hat{p}\cdot{\bf q})^2\dot{B}_q(t')\Big)\nonumber\\&&+\frac{3\omega_c(t')}{2}
\left\{\Phi(q,t')\,\Big(\delta_{ij}-\hat{p}_i\hat{p}_j\Big)+
\frac{1}{2}\Pi(q,t')\,\Big(\hat{q}_i-\hat{p}_i(\hat{p}\cdot\hat{q})\Big)\,\Big(\hat{q}_j-\hat{p}_j(\hat{p}\cdot\hat{q})\Big)\right\}\nonumber\\&&~~+\frac{3i}{2a(t')}\Big(
\delta_{ij}-\hat{p}_i\hat{p}_j\Big)\,(\hat{p}\cdot\hat{q})\int_{t_1}^{t'}dt'' \,\frac{\omega_c(t'')a(t'')}{R(t'')}\Omega(q,t'')\,\exp\left(-\int_{t''}^{t'}dt'''\frac{\omega_c(t''')}{R(t''')}\right)\nonumber\\&&+2\omega_c(t')\,\Big(\delta_{ij}-\hat{p}_i\hat{p}_j\Big)\,\hat{p}_k \delta u_k({\bf q},t_1)\Bigg]+2\Big(\delta_{ij}-\hat{p}_i\hat{p}_j\Big)\,\left[\frac{\delta T({\bf q},t_1)}{\bar{T}(t_1)}+\hat{p}_k\delta u_k({\bf q},t_1)\right]\;.\nonumber\\&&{}
\end{eqnarray}

Conventionally, the intensity matrix perturbation for scalar modes is written
\begin{eqnarray}
&&J_{ij}({\bf q},\hat{p},t)= \beta({\bf q})\Bigg\{\frac{1}{2}\Big(\Delta_T^{(S)}(q,\hat{q}\cdot\hat{p},t)-\Delta_P^{(S)}(q,\hat{q}\cdot\hat{p},t)\Big)\,\Big(\delta_{ij}-\hat{p}_i\hat{p}_j\Big)\nonumber\\&&~~~~~~+
\Delta_P^{(S)}(q,\hat{q}\cdot\hat{p},t)\Big)\,\left[\frac{\Big(\hat{q}_i-(\hat{q}\cdot\hat{p})\hat{p}_i\Big)\,
\Big(\hat{q}_j-(\hat{q}\cdot\hat{p})\hat{p}_j\Big)}{1-(\hat{p}\cdot\hat{q})^2}\right]\Bigg\}\;,
\end{eqnarray}
with $\Delta_T^{(S)}$ and $\Delta_P^{(S)}$ satisfying the Boltzmann equations
\begin{equation}
\dot{\Delta}_P^{(S)}(q,\mu,t)+i\left(\frac{q\mu}{a(t)}\right)\Delta_P^{(S)}(q,\mu,t)=-\omega_c(t)\Delta_P^{(S)}(q,\mu,t)+\frac{3}{4}\omega_c(t)\,(1-\mu^2)\Pi(q,t)\;,
\end{equation}
\begin{eqnarray}
&&\dot{\Delta}_T^{(S)}(q,\mu,t)+i\left(\frac{q\mu}{a(t)}\right)\Delta_T^{(S)}(q,\mu,t)=-\omega_c(t)\Delta_T^{(S)}(q,\mu,t)-2\dot{A}(q,t)+2q^2\mu^2\dot{B}(q,t)\nonumber\\&&~~~~~~+3\omega_c(t)\,\Phi(q,t)+
\frac{3}{4}\omega_c(t)(1-\mu^2)\Pi(q,t)+4\omega_c(t)\hat{p}_i\delta u_i;.
\end{eqnarray}
By expanding $\Delta_T^{(S)}$ and $\Delta_P^{(S)}$ in partial wave amplitudes
\begin{eqnarray}
\Delta_T^{(S)}(q,\mu,t)&=&\sum_{\ell=0}^\infty i^{-\ell}(2\ell+1)\,P_\ell(\mu)\,\Delta_{T\ell}^{(S)}(q,t)\,\\
\Delta_P^{(S)}(q,\mu,t)&=&\sum_{\ell=0}^\infty i^{-\ell}(2\ell+1)\,P_\ell(\mu)\,\Delta_{T\ell}^{(S)}(q,t)\;,
\end{eqnarray}
and integrating Eq.~(37) over the directions of $\hat{p}$, one obtains well known expressions for the source functions in terms of the partial wave amplitudes with $\ell\leq 2$:
\begin{eqnarray}
\Phi&=&\frac{1}{6}\Big[2\Delta^{(S)}_{T0}-\Delta^{(S)}_{P0}-\Delta^{(S)}_{T2}-\Delta^{(S)}_{P2}\Big]\\
\Pi&=&\Delta^{(S)}_{P0}+\Delta^{(S)}_{T2}+\Delta^{(S)}_{P2}\\
\Omega&=&\Delta^{(S)}_{T1}
\end{eqnarray}

Instead, we suggest the alternative of deriving coupled integral equations for $\Phi$, $\Pi$, and $\Omega$ by analytically integrating over $\hat{p}$ in Eq.~(36).  This gives
\begin{eqnarray}
&&\Phi(q,t)=\int_{t_1}^t dt'\;\exp\left(-\int_{t'}^t dt''\;\omega_c(t'')\right)\nonumber\\&&\times\Bigg[\dot{A}_q(t')F_1\left(q\int_{t'}^t\frac{dt''}{a(t'')}\right)
+q^2\dot{B}_q(t')F_2\left(q\int_{t'}^t\frac{dt''}{a(t'')}\right)\nonumber\\&&
+\frac{3\omega_c(t')}{2}\Big\{-\Phi(q,t')F_1\left(q\int_{t'}^t\frac{dt''}{a(t'')}\right)
+\frac{1}{2}\Pi(q,t')F_3\left(q\int_{t'}^t\frac{dt''}{a(t'')}\right)\Big\}\nonumber\\&&
+\frac{3}{2a(t')}F_4\left(q\int_{t'}^t\frac{dt''}{a(t'')}\right)\int_{t_1}^{t'}dt''\;\frac{\omega_c(t'')a(t'')}{R(t'')}\Omega(q,t'')\exp\left(-\int_{t''}^{t'}\frac{\omega_c(t''')\,dt'''}{R(t''')}\right)\Bigg]\;,\nonumber\\&&{}\\&& \Pi(q,t)=\int_{t_1}^t dt'\;\exp\left(-\int_{t'}^t dt''\;\omega_c(t'')\right)\nonumber\\&&\times\Bigg[-2\dot{A}_q(q,t')j_2\left(q\int_{t'}^t\frac{dt''}{a(t'')}\right)
+2q^2\dot{B}_q(t')F_5\left(q\int_{t'}^t\frac{dt''}{a(t'')}\right)\nonumber\\&&
+3\omega_c(t')\Big\{\Phi(q,t')j_2\left(q\int_{t'}^t\frac{dt''}{a(t'')}\right)
+\Pi(q,t')F_6\left(q\int_{t'}^t\frac{dt''}{a(t'')}\right)\Big\}\nonumber\\&&
+\frac{3}{a(t')}F_7\left(q\int_{t'}^t\frac{dt''}{a(t'')}\right)\int_{t_1}^{t'}dt''\;\frac{\omega_c(t'')a(t'')}{R(t'')}\Omega(q,t'')\exp\left(-\int_{t''}^{t'}\frac{\omega_c(t''')\,dt'''}{R(t''')}\right)\Bigg]\;,\nonumber\\&&{}\\&&\Omega(q,t)=\int_{t_1}^t dt'\;\exp\left(-\int_{t'}^t dt''\;\omega_c(t'')\right)\nonumber\\&&\times\Bigg[2\dot{A}_q(t')j_1\left(q\int_{t'}^t\frac{dt''}{a(t'')}\right)
-2q^2\dot{B}_q(t')F_{8}\left(q\int_{t'}^t\frac{dt''}{a(t'')}\right)\nonumber\\&&
+\frac{3\omega_c(t')}{2}\Big\{-2\Phi(q,t')j_1\left(q\int_{t'}^t\frac{dt''}{a(t'')}\right)
+\frac{1}{2}\Pi(q,t')F_{9}\left(q\int_{t'}^t\frac{dt''}{a(t'')}\right)\Big\}\nonumber\\&&
+\frac{3}{2a(t')}F_{10}\left(q\int_{t'}^t\frac{dt''}{a(t'')}\right)\int_{t_1}^{t'}dt''\;\frac{\omega_c(t'')a(t'')}{R(t'')}
\Omega(q,t'')\exp\left(-\int_{t''}^{t'}\frac{\omega_c(t''')\,dt'''}{R(t''')}\right)\Bigg]\;,\nonumber\\&&{}
\end{eqnarray}
where 
\begin{eqnarray}
F_1(v)&\equiv& j_1(v)/v-j_0(v)\;,\\
F_2(v)&\equiv&j_1(v)/v-j_2(v)-j_2(v)/v^2+j_3(v)/v(v)\;,\\
F_3(v)&\equiv&j_1(v)/v+ j_2(v)/v^2-j_3(v)/v\;,\\
F_4(v)&\equiv& j_1(v)-j_2(v)/v\;,\\
F_5(v)&\equiv&-2j_2(v)/v^2+5j_3(v)-j_4(v)\;,\\
F_6(v)&\equiv&j_0(v)-2j_1(v)/v+2j_2(v)+2j_2(v)/v^2-5j_3(v)/v+j_4(v)\;,\nonumber\\&&{}\\
F_7(v)&\equiv&-2j_2(v)/v+j_3(v)\;,\\
F_8(v)&\equiv&3j_2(v)/v-j_3(v)\;,\\
F_9(v)&\equiv&-j_1(v)+3j_2(v)/v-j_3(v)\;,\\
F_{10}(v)&\equiv&2j_1(v)/v-2j_2(v)\;,
\end{eqnarray}
(We have dropped the terms involving the plasma temperature and velocity perturbations at the initial time $t_1$, because as long as the collision rate is much larger than the expansion rate the photon distribution is given in terms of $\delta T/\bar{T}$ and $\delta u$ by an equilibrium formula like Eq.~(14), and as long as the perturbation is outside the horizon  $\delta T/\bar{T}$ and $\delta u$ grow in the adiabatic mode, so that if $t_1$ is taken sufficiently early in the era when the collision rate is large {\em and} the perturbation is outside the horizon, the initial-data terms become negligible.)
One may again expect that these coupled integral equations may be solved by iteration, as done for the tensor modes.

\begin{center}
{\bf III.  Neutrinos}
\end{center}

In the absence of collisions, the Boltzmann equation for either massless or massive neutrinos for the metric (1) reads
\begin{equation}
\frac{\partial n}{\partial t}+\frac{p^i}{p^0}\frac{\partial n}{\partial x^i}=-\frac{\partial n}{\partial p_i}\frac{p^jp^k}{p^0}\frac{\partial g_{jk}}{\partial x^i}\;.
\end{equation}
Here $n({\bf x},{\bf p},t)$ is the phase space density of neutrinos, regarded as a function of the components $x^i$, $p_i$ and the time $t$, while $p^i$ and $p^0$ are functions of the $x^i$ and $t$ as well as the $p_i$, given by $p^i=g^{ij}p_j$ and $p^0=\sqrt{g^{ij}p_ip_j+m^2}$.  (The derivation of Eq.~(58) is given in Ref. [4] for massless neutrinos, but precisely the same derivation applies also for massive neutrinos.)  For weak perturbations, we write
\begin{equation}
n({\bf x},{\bf p},t)=\bar{n}_\nu\Big(a(t)\,\sqrt{g^{ij}({\bf x},t)p_ip_i}\Big)+\delta n({\bf x},{\bf p},t)\;,
\end{equation}
where $\bar{n}_\nu$ is a time-independent function of its argument
\begin{equation}
\bar{n}_\nu(p)\equiv \frac{1}{(2\pi)^3}\left[\exp\left(\frac{\sqrt{p^2+a^2(t_1)m^2}}{k_{\cal B}a(t)T(t)}\right)+1\right]^{-1}\;.
\end{equation}
Here  $\delta n$ is a small correction, representing the dynamical rather than purely geometric effect of metric perturbations on the neutrinos, and $t_1$ is  the time the neutrinos went out of equilibrium with the baryonic plasma.  (Probably all neutrinos have masses much less than $k_{\rm B}T(t_1)$, in which case the term $a^2(t_1)m^2$ in the square root can be neglected.)  Expanding to first order in $\delta n$ and $\delta g_{ij}=a^2h_{ij}$, after many cancellations this gives
\begin{equation}
\frac{\partial \delta n}{\partial t}+\frac{\partial \delta n}{\partial x^i}\frac{p_i}{a\sqrt{p_kp_k+a^2m^2}}=\frac{\bar{n}'_\nu(\sqrt{p_kp_k})}{2\sqrt{p_kp_k}}\dot{h}_{kl}p_kp_l\;.
\end{equation}
We again write $h_{ij}$ as a Fourier integral (8), and seek a solution in the form
\begin{equation}
\delta n({\bf x},{\bf p},t)=\int d^3q\;e^{i{\bf q}\cdot{\bf x}}\delta n({\bf q},{\bf p},t)\;.
\end{equation}
Then Eq.~(61) becomes an ordinary differential equation, with $\partial/\partial x^i$ replaced with $iq_i$.  
Instead of solving this equation numerically with a truncated partial wave expansion, it can be solved analytically, as a line of sight integral
\begin{eqnarray}
\delta n({\bf q},{\bf p},t)&=&\frac{\bar{n}'_\nu(\sqrt{p_kp_k})}{2\sqrt{p_kp_k}}\int_{t_1}^t dt'\;\exp\left(-i{\bf q}\cdot{\bf p}\int_{t'}^t\frac{dt''}{a(t'')\sqrt{p_kp_k+a^2(t'')m^2}}\right)\nonumber\\&&\times\;\dot{h}_{kl}({\bf q},t')\,p_kp_l\;.
\end{eqnarray}
where $t_1$ is any time taken early enough so that $\delta n({\bf x},{\bf p},t_1)$ is negligible. 

For the foreseeable future, the neutrino distribution will be important only in calculating the neutrino contribution to the energy-momentum tensor.  The first-order perturbation to the contribution of each species of neutrino or antineutrino to the mixed components of the energy-momentum tensor takes the simple form
\begin{eqnarray}
\delta T_\nu^i{}_j({\bf x},t)&=&\frac{1}{a^4(t)}\int\left(\prod_{k=1}^3 dp_k\right) \delta n_\nu({\bf x},{\bf p},t)\,\frac{p_ip_j}{\sqrt{p_kp_k+a^2(t)m^2}}\;,\\
\delta T_\nu^0{}_j({\bf x},t)&=&\frac{1}{a^3(t)}\int\left(\prod_{k=1}^3 dp_k\right) \delta n_\nu({\bf x},{\bf p},t)\,p_j\;,\\
\delta T_\nu^0{}_0({\bf x},t)&=&-\frac{1}{a^4(t)}\int\left(\prod_{k=1}^3 dp_k\right) \delta n_\nu({\bf x},{\bf p},t)\,\sqrt{p_kp_k+a^2(t)m^2}\;,
\end{eqnarray}
with all other first-order contributions nicely cancelling.  In the tensor mode,  for which $q_ih_{ij}=h_{ii}=0$, the only non-vanishing perturbations are to the space-space components of the energy-momentum tensor, which are needed in calculating the viscous damping of gravitational waves[4].  These take the form
\begin{eqnarray}
&&\delta T_\nu^i{}_j({\bf x},t)=\int d^3q\;e^{i{\bf q}\cdot{\bf x}}\int_0^\infty 4\pi p^5\,\bar{n}_\nu'(p)\,dp\nonumber\\&&~~~~~~~~~\times \int_{t_1}^{t}dt'\;K\left(qp\int_{t'}^t \frac{dt''}{a(t'')\sqrt{p^2+a^2(t'')m^2}}\right)\,\frac{\dot{h}_{ij}({\bf q},t')}{\sqrt{p^2+a^2(t')m^2}}\;,\nonumber\\&&{}
\end{eqnarray}
where $K(v)\equiv j_2(v)/v^2$.
For scalar modes, we get contributions to all components, of the form
\begin{equation}
\delta T_\nu^i{}_j=\delta_{ij}\delta p_\nu+\partial_i\partial_j\pi_\nu\;,~~~\delta T_\nu^0{}_j=\frac{4}{3}\bar{\rho}_\nu\partial_j \delta u_\nu\;,~~~\delta T^0{}_0=-\delta\rho_\nu\;,
\end{equation}
in which $\delta p_\nu$, $\pi_\nu$, $\delta u_\nu$ and $\delta \rho_\nu$ may be identified as the pressure perturbation, scalar anisotropic inertia, velocity potential, and energy density perturbation, respectively, for a given neutrino species.
Inserting Eqs.~(32) and (33) in Eq.~(63) and then using the result in Eqs.~(64)--(66) gives
\begin{eqnarray}
&&\delta p_\nu({\bf x},t)=\frac{1}{2a^4(t)}\int d^3q\;e^{i{\bf q}\cdot {\bf x}}\beta({\bf q})\int_0^\infty \frac{4\pi p^5\, \bar{n}'_\nu(p)\,dp}{\sqrt{p^2+a^2(t)m^2}}\nonumber\\&&~~~\times\int_{t_1}^t dt'\;\left[\dot{A}_q(t')F_{11}\left(qp\int_{t'}^t \frac{dt''}{a(t'')\sqrt{p^2+a^2(t'')m^2}}\right)\right.\nonumber\\&&~~~~~~~~~\left.-\,q^2\dot{B}_q(t')F_{12}\left(qp\int_{t'}^t \frac{dt''}{a(t'')\sqrt{p^2+a^2(t'')m^2}}\right)\right]\;,\\
&&\pi_\nu({\bf x},t)=\frac{1}{2a^4(t)}\int d^3q\;e^{i{\bf q}\cdot {\bf x}}\beta({\bf q}) \int_0^\infty dp\;\frac{4\pi p^5\, \bar{n}'_\nu(p)}{\sqrt{p^2+a^2(t)m^2}}\nonumber\\&&~~~\times\int_{t_1}^t dt'\;\left[q^{-2}\dot{A}_q(t')j_2\left(qp\int_{t'}^t \frac{dt''}{a(t'')\sqrt{p^2+a^2(t'')m^2}}\right)\right.\nonumber\\&&~~~~~~~~~\left.-\,\dot{B}_q(t')F_5\left(qp\int_{t'}^t \frac{dt''}{a(t'')\sqrt{p^2+a^2(t'')m^2}}\right)\right]\;,\\
&&\frac{4}{3}\bar{\rho}_\nu(t)\delta u_\nu({\bf x},t)=-\frac{1}{2a^3(t)}\int d^3q\;e^{i{\bf q}\cdot {\bf x}}\beta({\bf q})\int_0^\infty 4\pi p^4 \bar{n}'_\nu(p)\,dp\nonumber\\&&~~~\times\int_{t_1}^t dt'\;\left[q^{-1}\dot{A}_q(t')j_1\left(qp\int_{t'}^t \frac{dt''}{a(t'')\sqrt{p^2+a^2(t'')m^2}}\right)\right.\nonumber\\&&~~~~~~~~~\left.-\,q\dot{B}_q(t')F_8\left(qp\int_{t'}^t \frac{dt''}{a(t'')\sqrt{p^2+a^2(t'')m^2}}\right)\right]\;,\\
&&\delta \rho_\nu({\bf x},t)=\frac{1}{2a^4(t)}\int d^3q\;e^{i{\bf q}\cdot {\bf p}}\beta({\bf q})\int_0^\infty 4\pi p^3\, \bar{n}'_\nu(p)\sqrt{p^2+a^2(t)m^2}\,dp\nonumber\\&&~~~\times\int_{t_1}^t dt'\;\left[\dot{A}_q(t')j_0\left(qp\int_{t'}^t \frac{dt''}{a(t'')\sqrt{p^2+a^2(t'')m^2}}\right)\right.\nonumber\\&&~~~~~~~~~\left.-\,q^2\dot{B}_q(t')F_{13}\left(qp\int_{t'}^t \frac{dt''}{a(t'')\sqrt{p^2+a^2(t'')m^2}}\right)\right]\,
\end{eqnarray}
where 
\begin{equation}
F_{11}(v)\equiv j_1(v)/v\;,~~F_{12}(v)=j_2(v)/v^2-j_3(v)/v\;,~~F_{13}(v)=j_1(v)/v-j_2(v)\;.
\end{equation}
It is only in the case $m=0$ that the integrals over neutrino energies can be done separately from the integrals over time, and give  results simply proportional to $\bar{\rho}_{\nu} $.

The contribution of photons to the perturbations in the energy momentum tensor can be calculated in a similar way, by using Eqs.~(64)--(66) with $a^2\delta n^{ii}$ in place of $\delta n$ (and of course with $m=0$), taking the integral of $\delta n^{ii}$ over photon energies from the line of sight integrals (28) or(36) for tensor or scalar modes, respectively.

\vspace{12pt}

{\em Added Note:}  After the preprint of this paper was first circulated, I learned that similar suggestions regarding the calculation of the photon source terms have been made in the preprint of a paper by D. Beskaran, L. P. Grischchuk, and R. G. Polnarev, gr-qc/0605100.  Their equation (61) is the same integral equation for the source terms of tensor perturbations to the cosmic microwave background as presented here in equation (29).  However, for scalar modes they do not give an integral equation for the baryonic plasma streaming velocity, and so instead of  the three coupled integral equations found here, they give  two coupled integral equations, in which the plasma velocity as well as the gravitational field perturbations appear as inputs.

\vspace{12pt}

I am grateful to Eiichiro Komatsu for frequent helpful conversations, and to Raphael Flauger for preparing the figure and pointing out some typographical errors.  This material is based upon work supported by the National Science Foundation under Grant Nos. PHY-0071512 and PHY-0455649 and with support from The Robert A. Welch Foundation, Grant No. F-0014.

\begin{center}
{\bf References}
\end{center}

\begin{enumerate}
\item P.J.E. Peebles and J. T. Yu, Astrophys. J. {\bf 162}, 815 (1970); R. A. Sunyaev and Ya. B. Zel'dovich, Astrophys. Space Sci. {\bf 7}, 3 (1970).
\item M. L. Wilson and J. Silk, Astrophys. J. {\bf 243}, 14 (1981); J. R. Bond and G. Efstathiou, Astrophys. J. {\bf 285}, L45 (1984); R. Crittenden, J. R. Bond, R. L. Davis, G. Efstathiou, and P. J. Steinhardt, Phys. Rev. Lett. {\bf 71}, 324 (1993); C.-P. Ma and E. Bertschinger, Astrophys. J. {\bf 455}, 7 (1995).
\item   U. Seljak and M. Zaldarriaga, Astrophys. J. {\bf 469}, 437 (1996).
\item S. Weinberg,  Phys. Rev. D {\bf 69}. 023501 (2004).
\item R. Crittenden {\em et al.}, ref. [2].

\end{enumerate}

\end{document}